\documentclass[%
 reprint,
 amsmath,amssymb,
 aps,
]{revtex4-2}

\usepackage{graphicx}
\usepackage{dcolumn}
\usepackage{bm}
\usepackage{xcolor}
\usepackage{qtree}
\usepackage{tikz-cd}
\usepackage{multirow}
\usepackage[
        colorlinks=true,
        linkcolor=black,
        urlcolor=blue,
        citecolor=blue
    ]{hyperref}
\usepackage{url}

\newcommand{\chemform}{EuCd$_2$As$_2$}

\usepackage{ulem} 

\begin{document}

\preprint{APS/123-QED}

\title{\texorpdfstring{
Influence of magnetism, strain and pressure on the band topology of EuCd$_2$As$_2$
}{}}

\author{Adrian Valadkhani${}^{1}$}
\email{valad@itp.uni-frankfurt.de}
\author{Mikel Iraola${}^{2}$}
\author{Axel Fünfhaus${}^{1}$}
\author{Young-Joon Song${}^{1}$}
\author{Libor Šmejkal${}^{3,4}$}
\author{Jairo Sinova${}^{3,4}$}
\author{Roser Valentí${}^{1}$}%
 \email{valenti@itp.uni-frankfurt.de}
 \affiliation{%
 ${}^{1}$Institut für Theoretische Physik, Goethe-Universität Frankfurt, Max-von-Laue-Strasse 1, 60438 Frankfurt am Main, Germany\looseness=-1
}
\affiliation{%
 ${}^{2}$Donostia International Physics Center, 20018 Donostia-San Sebastian, Spain\looseness=-1
}%
\affiliation{%
 ${}^{3}$Institute for Physics, Johannes Gutenberg-University, 55122 Mainz, Germany\looseness=-1
}%
\affiliation{%
 ${}^{4}$Institute of Physics, Czech Academy of Sciences, Cukrovarnická 10, 162 00 Praha 6, Czech Republic\looseness=-1
}%

\date{\today}

\begin{abstract}
Motivated by the wealth of proposals and realizations of nontrivial topological phases in \chemform, such as a Weyl semimetallic state and the recently controversially discussed semimetallic versus semiconductor behavior in this system, we analyze in this work the role of the delicate interplay of Eu magnetism, strain and pressure on the realization of such phases. For that we invoke a combination of a group theoretical analysis with {\it ab initio} density functional theory calculations and uncover a rich phase diagram with various non-trivial topological phases beyond a Weyl semimetallic state, such as axion and topological crystalline insulating phases, and discuss their realization.
\end{abstract}

\maketitle

\section{\label{sec:intro}Introduction}

Trigonal ACd$_2$X$_2$ pnictides (A= lanthanoid or alkaline earth metal, X = P, As, Sb) ~\cite{zaac.19966220418,zaac.201100179,C8DT02955E,doi:10.1063/1.4961565} have lately received a lot of attention due to the presence of non-trivial topology associated
to their electronic properties. In particular,
EuCd$_2$As$_2$ has been intensively discussed as a possible Weyl semimetal candidate~\cite{PhysRevB.99.245147,PhysRevB.100.201102,yu2022pressure}, although the semimetallic 
behavior itself at ambient pressure is controversially being debated~\cite{santos2023}.

\begin{figure}[ht]
 	\centering
 	\includegraphics[width=1.0\linewidth]{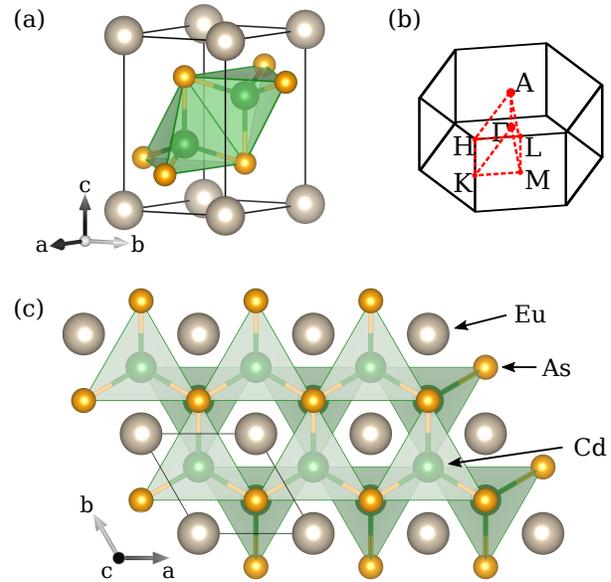}
 	\caption{(a) Primitive unit cell of \chemform. (b) First Brillouin zone (black) of the $P\bar{3}m1$ trigonal space group with all high symmetry lines and points shown in red. The labels indicate the high symmetry points. (c) Top view of the crystal structure with additional unit cells showing the honeycomb structure of CdAs$_{4}$ tetrahedra.} 
 	\label{fig:structure}
 \end{figure}

This compound consists of stacked layers
of edge-shared CdAs$_4$ tetrahedra separated by hexagonal layers of Eu ions
(Fig.~\ref{fig:structure}).
At ambient pressure the Eu$^{2+}$ network orders at $T_N\approx9.5${ }K in
an A-type antiferromagnetic pattern where the Eu moments align  
ferromagnetically (FM) in the $ab$ plane and antiferromagnetically (AFM) along the $c$-direction with
the N\'eel vector pointing in the $ab$ plane~\cite{PhysRevB.94.045112,PhysRevB.97.214422}.

In recent years, a few theoretical studies based on density functional theory have predicted the possibility of topological states in this material. 
The authors of Ref.~\cite{PhysRevB.98.201116} suggested the A-type AFM ordered \chemform{} with Eu moments pointing along the $c$-direction (out-of-plane) as a candidate for hosting a pair of Dirac points protected by inversion $\mathcal{P}$, time reversal  $\mathcal{T}$ symmetry combined with a $ \tau\,=\,(0,0,1/2)^T $ translation of the magnetic unit cell and $C_3$ rotational symmetry. 
Interestingly, the predicted antiferromagnetic Dirac fermions belong to a different type than, for instance, in \mbox{CuMnAs}~\cite{tang2016,vsmejkal2017} where the Dirac fermions are protected by a $\mathcal{PT}$ symmetry and a nonsymmophic screw rotation.
In \mbox{CuMnAs} a tuning of crystallographic and magnetic symmetries can result, for instance, in enhanced transport properties~\cite{vsmejkal2017}. 
Such a tunability in \chemform{} is presently being intensively scrutinized.
For example, it has been proposed~\cite{PhysRevB.99.245147} that doping with Ba a fully ferromagnetically  ordered EuCd$_2$As$_2$ with Eu moments pointing out-of-plane, would turn the system into a Weyl semimetal~\cite{armitage2018,heinsdorf2021}.
The presence of antiferromagnetic or ferromagnetic order was recently reported~\cite{PhysRevB.101.140402} as
a function of band filling, and a stable ferromagnetic order was observed with
Ba doping~\cite{taddei2020spincanting}, albeit with an additional canting of the Eu magnetization, which sensitively affects the realization of Weyl physics.

 EuCd$_2$As$_2$ appears to represent a convenient system for studying phase transitions by application of 
 pressure or strain~\cite{li2022} as well. Efforts to investigate this material under pressure~\cite{PhysRevB.104.155124} 
found a transition from an A-type AFM order to a FM order  at a pressure
of $2\,$GPa with in-plane oriented Eu moments. The long-sought Weyl semimetal 
was predicted to appear at much higher pressures along with
a second transition from FM 
order with in-plane Eu moments to FM order with out-of-plane
Eu moments. However, we became aware of a recent publication~\cite{Greeshma2023} that sees no transition to the FM out-of-plane order up to $\sim40$\,GPa.
Furthermore, recent theoretical work~\cite{Cuono2023} suggests that \chemform{} is more of a trivial semiconductor than a topologically nontrivial semimetal.

Despite this wealth of theoretical predictions and experimental results, the question remains whether further topological phases exist or at least can be accessed by manipulating certain parameters. 
In this work we invoke a combination of a group theoretical analysis in the context of topological quantum chemistry~\cite{elcoro2020magnetic, Bradlyn2017, Po2017, Cano2021,Kurthoff2017} and \textit{ab initio} density functional theory. With these methods we explore systematically the topology and electronic properties of \chemform{} as a function of Eu magnetism, strain and pressure.
The trends of these predictions provide important insights into how to tune this or related systems in and out of topologically nontrivial phases.

Depending on the type of magnetic ordering of Eu moments, strain and pressure, we find diverse non-trivial topological phases beyond Weyl semimetals including axion insulators~\cite{PhysRevB.98.245117} and topological crystalline insulators. 

The paper is organized as follows.
In Section~\ref{Computational}  we describe the computational methods used for our study. 
In Section~\ref{sec:results} we present the results of the topology of the electronic structure of \chemform{}, starting with the analysis of uniaxial and shear strain effects on non-magnetic \chemform{} (Section~\ref{sec:nm}). We proceed in Section~\ref{sec:afm1} with the role of A-type antiferromagnetic ordering on the topology of \chemform{} and in Section~\ref{sec:afm2} we analyze the effect of pressure and strain on this particular magnetic phase. Sections~\ref{sec:fm1}, \ref{sec:fm2} are dedicated to ferromagnetic \chemform{}. 
In Section~\ref{sec:U} we investigate how correlations as described by the Hubbard $U$ affect the topology of \chemform{}.   
Finally in Section~\ref{sec:concldisc} we present our conclusions.

\section{\label{Computational} Computational methods}
We perform electronic structure calculations within density functional theory (DFT) by using
 two different basis sets: the pseudo-potential augmented plane-wave Vienna Ab initio Simulation Package (VASP)~\cite{PhysRevB.47.558} and the full-potential (linearized) augmented plane-wave + local orbitals method WIEN2k~\cite{Blaha2020}.
 All calculations were performed with the Perdew–Burke–Ernzerhof (PBE) generalized gradient approximation (GGA)~\cite{PhysRevLett.77.3865}, and a plane-wave cutoff of 500\,eV. For the non-magnetic  and ferromagnetic cases, which have the same primitive unit cell, an $ 18\times18\times9 $ $ k $-point mesh was utilized. A different $k$-mesh of size $ 22\times22\times6 $ was chosen for the antiferromagnetic case. 
 Depending on the calculation, we also considered spin polarization, spin-orbit coupling (SOC) and a Hubbard $U = 5$\,eV interaction, which localizes the Eu 4f bands closely to the positions visible from experiments~\cite{Ma2019,Ma2020}.

The crystal information for EuCd$_2$As$_2$ was taken from Ref.~\cite{zaac.201100179}.  
We relaxed the structure with VASP (GGA+SOC+U)  until forces smaller than $0.01\,$eV/\AA~ 
were reached.
On the one hand, we considered
 A-type AFM order with
Eu moments pointing out-of-plane. 
The relaxed structure is in 
very good agreement with the experimental one (see Table in Appendix~\ref{app:gs:table}).  
On the other hand, we note that relaxations with a ferromagnetic initial configuration hardly change the results. 
Relaxations of the crystal structures under pressure or strain conditions were done with VASP.

For the group theoretical analysis we use  (M)vasp2trace~\cite{xu_high-throughput_2020,elcoro2020magnetic}.
As a cross check for the electronic structures and irreducible (co)representations, calculations
with WIEN2k were also performed.

\section{\label{sec:results} Results and discussion}
\subsection{Strain effects on the non-magnetic structure \label{sec:nm}}

We first investigate the effect of strain on the symmetry of non-magnetic 
\chemform{}. 

The arrangement of ions forms a trigonal structure with lattice parameters $a=b$ and $c$, and angles $\alpha=\beta=\pi/2, \gamma=2\pi/3$ between the primitive lattice vectors (see Fig. \ref{fig:structure} (a), (c)).
The symmetry of this crystal is given by the space group $P\bar{3}m1$ (No. 164, see Appendix~\ref{app:symmetry lowering}). 
Eu, Cd and As sit in Wyckoff positions $ 1a $, $2d$ and $ 2d $, respectively.

\begin{figure}[ht]
 	\centering
 	\includegraphics[width=1.0\linewidth]{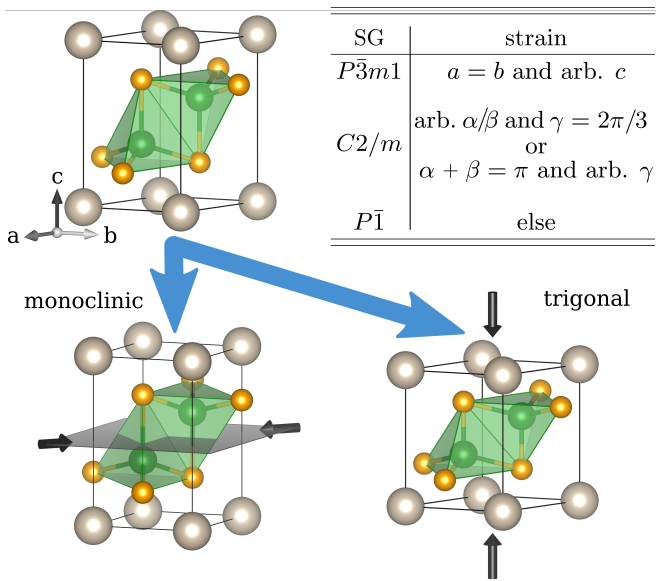}
 	\caption{
 	Schematic representation of the action of shear strain (lower left corner)
 	and uniaxial strain (lower right corner) on the unit cell of \chemform{} (upper left corner).
 	The table on the upper right corner lists the possible subgroups of $P\bar{3}m1$
 	that can be achieved applying strain (see main text) and the conditions imposed on the
 	lattice parameters $a, b, c$ and the lattice angles $\alpha, \beta, \gamma$.
 	}
 	\label{fig:uniaxshear}
 \end{figure}

\begin{figure*}[!ht]
		\centering
		\includegraphics[width=0.8\textwidth] {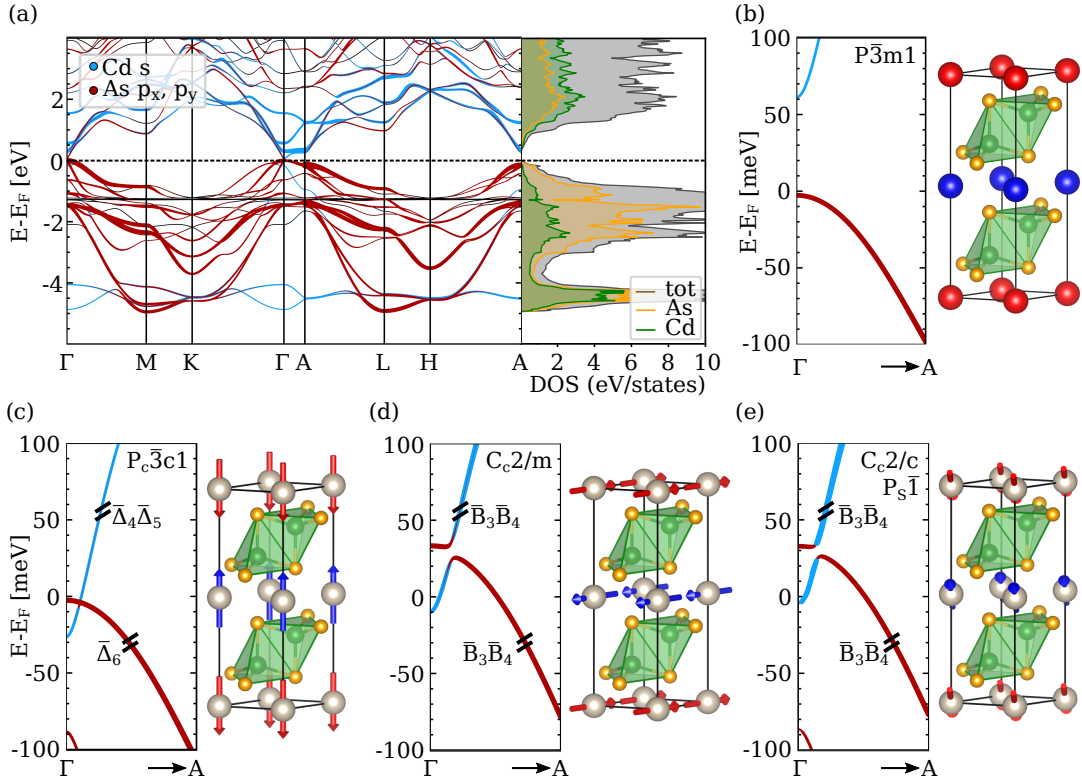}
		\caption{
        Electronic structure and unit cells of \chemform{} for A-type antiferromagnetic orders.
        (a) GGA+U band structure and DOS obtained with $U=5$\,eV, and without SOC.
        (b) Blow-up of the electronic structure near the Fermi surface along the $\Gamma$-A path. The magnetic unit cell is shown on the right. The {red} and {blue} atoms denote Eu with up and down spin initialization respectively.
        (c-e) GGA+U+SOC band structures of  \chemform{} along the $\Gamma$-A path obtained for moments aligned parallel to (c) the principal axis, $\textbf{M} || 3^{+}_{001}$, (d) a 2-fold rotation axis, e.g. in [100] direction, $\textbf{M} || 2_{100}$, (e) a mirror plane, $\textbf{M} || m_{100}$.
        The insets of (b-e) denote the magnetic space groups (top) and the corresponding irreducible coreps (their dimension is indicated by the black lines on the bands).
        } 
	    \label{fig:EuCd2As2:afm:dft}
\end{figure*}

In what follows we analyze the possible subgroups of $P\bar{3}m1$ that one can achieve by applying strain, for further discussion see Appendix~\ref{app:symmetry lowering}.
We consider distortions which change (i) the length of the lattice vectors -- known as {\it axial} strain -- and (ii) the angle between the lattice vectors, i.e. {\it shear} strain (see Fig.~\ref{fig:uniaxshear}).
Uniaxial strain along the $c$-direction (Fig.~\ref{fig:uniaxshear} lower right corner), its corresponding analog with equal strain on $a$- and $b$-directions -- known as biaxial (or epitaxial) strain -- and the combination of both, preserve the trigonal symmetry. Any other uniaxial strain perturbation will drive a transition to a space group of lower symmetry.
For instance, uniaxial strain along either $ a$ or $b$ direction breaks every symmetry except for inversion and translations, reducing the symmetry to $P\bar{1}$ (No. 2).

Shear strain (Fig.~\ref{fig:uniaxshear} lower left corner) is 
able to preserve one of the three secondary two-fold rotation axes of the trigonal lattice system, and the reflection plane perpendicular to the rotation axis, if either $\alpha$ or $\beta$ is changed with $\gamma$ fixed.
The symmetry is lowered to the space group $C2/m$ (No. 12) in this case.
Alternatively, changing both $\alpha$ and $\beta$ drives a transition to the same monoclinic space group, as long as their sum is kept constant at $\alpha +\beta = \pi$.
For the latter case $\gamma$ can take an arbitrary physically meaningful value. 
All other conceivably ways of applying shear strain reduce the spatial symmetry to $P\bar 1$, as is shown in the table in Fig.~\ref{fig:uniaxshear}.

In Section~\ref{sec:afm2} we analyze the effect of such strains on the topological properties of magnetic \chemform{}.

\begin{table*}[]
\begin{ruledtabular}\begin{tabular}{cccccc}
  \textbf{M} direction            &    space group    & \multicolumn{4}{c}{band corepresentation} \\ \hline
\multirow{4}{*}{$\textbf{M} ||3_{001}^{+}$}  & \multirow{4}{*}{$P_c \bar{3}c1$}  &     &  $({}^{1}\!\bar{E} \uparrow P_c \bar{3}c1)_{4d}$   &  $({}^{2}\!\bar{E} \uparrow P_c \bar{3}c1)_{4d}$   &  $(\bar{E} \uparrow P_c \bar{3}c1)_{4d}$  \\\cline{3-6}
&                                                               &$\Gamma$&   $\bar{\Gamma}_8(2) \oplus \bar{\Gamma}_9(2)$  &  $\bar{\Gamma}_8(2) \oplus \bar{\Gamma}_9(2)$   &  $\bar{\Gamma}_4\bar{\Gamma}_5(2) \oplus \bar{\Gamma}_6\bar{\Gamma}_7(2)$  \\
&                                                               &$\Delta$&   $2 \bar{\Delta}_6(2) $  &   $2 \bar{\Delta}_6(2) $  &   $2 \bar{\Delta}_4\bar{\Delta}_5(2) $ \\
&                                                               & A    &    $\bar{A}_5(2) \oplus \bar{A}_6(2)$  &  $\bar{A}_5(2) \oplus \bar{A}_6(2)$   &  $ 2 \bar{A}_4 (2)$   \\ \hline
\multirow{4}{*}{$\textbf{M} ||2_{100}$} & \multirow{4}{*}{$ C_{c}2/m $} &    &  $({}^{1}\!\bar{E} \uparrow C_{c}2/m)_{8i}$   &   $({}^{2}\!\bar{E} \uparrow C_{c}2/m)_{8i}$  &    \\\cline{3-6}
&                                                               &$\Gamma$&  $2 \bar{\Gamma}_3\bar{\Gamma}_4(2) \oplus 2 \bar{\Gamma}_5\bar{\Gamma}_6(2)$   &  $2 \bar{\Gamma}_3\bar{\Gamma}_4(2) \oplus 2 \bar{\Gamma}_5\bar{\Gamma}_6(2)$   &    \\
&                                                               & B      &   $4 \bar{B}_3\bar{B}_4(2) $  &  $4 \bar{B}_3\bar{B}_4(2) $   &    \\
&                                                               &   A    &  $2 \bar{A}_3\bar{A}_5(2) \oplus 2 \bar{A}_4\bar{A}_6(2)$   &   $2 \bar{A}_3\bar{A}_5(2) \oplus 2 \bar{A}_4\bar{A}_6(2)$   &    \\ \hline
\multirow{4}{*}{$\textbf{M} ||m_{100}$ or else} & \multirow{4}{*}{$C_c2/c$ or $P_S \bar{1}$} &     & $(\bar{A} \uparrow C_c2/c)_{8i}$      &  $(\bar{A} \uparrow P_S \bar{1})_{4i}$   &    \\\cline{3-6}
&                                                                               &$\Gamma$&   $2 \bar{\Gamma}_3\bar{\Gamma}_4(2) \oplus 2 \bar{\Gamma}_5\bar{\Gamma}_6(2)$  &  $\bar{\Gamma}_2\bar{\Gamma}_2(2) \oplus \bar{\Gamma}_3\bar{\Gamma}_3(2)$   &    \\
&                                                                               &   B/GP   &   $4 \bar{B}_3\bar{B}_4(2) $  &  $2 \overline{GP}_2\overline{GP}_2(2)$   &    \\
&                                                                               &   A/Z   &   $ 4 \bar{A}_2 (2)$   &  $ 2 \bar{Z}_2\bar{Z}_3 (2)$   &   
\end{tabular}\end{ruledtabular}
\caption{Alignment of the Eu A-type antiferromagnetic vector along specific directions and their resulting space groups with the irreducible corepresentations for the given band corepresentation along the symmetry path between $\Gamma$ and A indicated. The notation was adapted from~\cite{Campbell2022}.}
\label{tab:afm:irreps}
\end{table*}

\subsection{\label{sec:afm1} A-type antiferromagnetic ground states}

 The arrangement of the Eu magnetic moments in specific directions may lead to a reduction of conserved symmetries and affects the topology of the electronic structure at the Fermi surface.
 Experimentally, in-plane A-type AFM order has been reported to be the preferred ground state~\cite{PhysRevB.101.140402,zaac.201100179} (for a discussion on the effect of the magnetic moments of Eu on Cd and As, see Appendix~\ref{app:afm:Ushift}).

In the A-type AFM states of \chemform{}, translation invariance by a primitive translation in $c$-direction $\tau$ is spontaneously broken due to the magnetic ordering, and only remains a symmetry in combination with time reversal $\mathcal{T}$ (see Appendix~\ref{app:symmetry lowering}). Therefore, the primitive unit cell of the AFM-structure is twice as large as in the non-magnetic structure (see Fig.~\ref{fig:EuCd2As2:afm:dft}~(b)). Nevertheless, $\mathcal{T}'=\mathcal{T}\tau$ symmetry can be seen as an effective time-reversal symmetry, since its combination with inversion symmetry $\mathcal{P}$ satisfies $(\mathcal{P}\mathcal{T}')^2=-1$ and protects Kramers degeneracy~\cite{kramers1934,PhysRevB.98.201116}.\\

In Fig.~\ref{fig:EuCd2As2:afm:dft} (a) we show the atom-resolved electronic structure and density of states of EuCd$_2$As$_2$ with Eu$^{2+}$ moments in the A-type AFM configuration as obtained from GGA+U ($U$ = 5\,eV) without including SOC, i.e. spin space and position space can be treated as decoupled from symmetries' perspective~\cite{Smejkal2022,Corticelli2022,Liu2022}. Cd $s$ and As $p_{x}, p_{y} $ orbitals constitute the dominant contribution at the Fermi level. The density of states near the Fermi level is concentrated along the  $\Gamma$-A high symmetry line, also denoted as $\Delta=(0,0,k_z)$ with $0<k_z<\pi/c$.
Without SOC the electronic structure has a direct gap of $64$\,meV at $\Gamma$, and trivial valence bands (see Fig.~\ref{fig:EuCd2As2:afm:dft} (b)).
  
With SOC, the direction of the A-type antiferromagnetic vector of Eu$^{2+}$ moments $\textbf{M}$ is essential to determine the nature
of the possible topological states. 
In particular, when $\textbf{M}$ is aligned along the principal axis (case denoted as $\textbf{M} ||3^{+}_{001}$), \chemform{}
hosts a pair of Dirac points in the vicinity of the Fermi level along the line $\Delta$ (see Fig.~\ref{fig:EuCd2As2:afm:dft}~(c)).  

As soon as the antiferromagnetic vector is not aligned with the principal axis (Fig.~\ref{fig:EuCd2As2:afm:dft} (d), (e)), the $3^{+}_{001}$ symmetry is broken.
It turns out that the possible in-plane magnetic groups are $ C_{c}2/m $, $C_{c}2/c$ or $P_{S}\bar{1}$~\cite{Campbell2022} (see Fig.~\ref{fig:EuCd2As2:afm:dft} (c)-(e)). 
In the face of this symmetry lowering, the high-symmetry line $\Delta$ -- parameterized by $(0,0,k_z)$ with $0<k_z<\pi/c$ -- stays invariant under the same symmetry operations as any point $(k_x,0,k_z)$ in the plane B -- with $0<k_x<\pi/a$.
As a consequence, we replace the line $\Delta$ by the plane B in Tab.~\ref{tab:afm:irreps}.

In the following, we discuss the potential of each magnetic structure with broken trigonal symmetry to host band crossings in the line $\Delta$ or plane B.
For that, we analyze the corepresentations (coreps) of \textit{ab initio} bands that could be involved in these crossings.
These irreducible coreps are written in Tab.~\ref{tab:afm:irreps}, together with the corresponding band corepresentations (for a short introduction of band (co)representations and basic group theoretical notions see Appendix \ref{app:bandrep}).

When collinear Eu$^{2+} $ moments are aligned with one of the secondary axes hosting a two-fold rotation symmetry (e.g. $2_{100}$, which is denoted as $\textbf{M} ||2_{100}$), the space group symmetry is given by the monoclinic magnetic group $C_{c}2/m$ (No. 12.63) (see Fig.~\ref{fig:EuCd2As2:afm:dft} (d)).
Cd and As ions sit within the (conventional base-centered) unit cell in WPs $8i$, whose sites are left invariant by site-symmetry groups isomorphic to the reflection point group $m$.
$s$-orbitals of Cd transform as the composite band corepresentation $[({}^{1}\!\bar{E} \oplus {}^{2}\!\bar{E})\!\uparrow\!G]_{8i}$.
$p_x$ and $p_y$-orbitals of As also transform separately as the same band corepresentation.
Furthermore, the band corepresentation $[({}^{1}\!\bar{E} \oplus {}^{2}\!\bar{E})\!\uparrow\!G]_{8i}$ has a single corep of the little group for points on the plane B.
Therefore, all bands induced from these orbitals transform as the same corep on this plane, which prevents the existence of symmetry protected band crossings.
Equivalently, the space group $C_{c}2/m$ lacks the symmetries needed to protect the Dirac points present in the trigonal-symmetry setting.

If collinear Eu$^{2+}$ moments are parallel to a mirror plane (case that we denote $ \textbf{M} ||m_{100}$), the magnetic space group of the crystal is $C_{c}2/c$ (No. 15.90).
The unit cell of this structure is shown in Fig.~\ref{fig:EuCd2As2:afm:dft} (e).
The application of shear strain may but does not have to lead to the loss of symmetry.
Cd and As ions are located within the unit cell in WPs $8i$, on sites whose site-symmetry groups are isomorphic to the point group $m'$.
Since the unitary subgroup of this group is the identity $1$, all Cd $s$ and As $p$-orbitals transform separately as the same corep $2 \bar A$ of $m'$. 
As a consequence, all bands induced from these orbitals transform on the points of the plane B as the same corep $\bar B_3 \bar B_4$, which forbids the existence of symmetry protected band crossings there.

Any other A-type AFM alignment of magnetic moments reduces the symmetry to the magnetic space group $P_{S}\bar{1}$ (No 2.7) (see Fig.~\ref{fig:EuCd2As2:afm:dft}~(e)). 
Regardless of which strain or stress type is applied, this subgroup remains unaltered.
The Wyckoff positions of Cd and As are $4i$, and the site-symmetry groups are isomorphic to the identity group $1$.
Thus, irreducible coreps on the plane B of the band corepresentations corresponding to Cd $s$ and As $p$-orbitals are of the same form as in $C_{c}2/c$, as it is written in the fourth row of Tab.~\ref{tab:afm:irreps}.
This implies that crossings between bands do not occur on this plane, so that the band structure does not exhibit Dirac points.
Additionally, an avoided crossing with inverted band characters occurs, as depicted in Fig.~\ref{fig:EuCd2As2:afm:dft}~(d-e).

The relevant topological index for $C_c 2/m$, $C_c 2/c$ and $P_S \bar{1}$ is $\eta_{4I}$~\cite{elcoro2020magnetic}, as time reversal symmetry is broken but inversion symmetry is preserved
\footnote{The density of states of the structures analyzed in figure \ref{fig:EuCd2As2:afm:dft}(c-e) have well seperated valence and conduction bands but are gapless, wherefore, an analysis using symmetry indicators can be done.}. For the A-type AFM bandstructures shown in Fig.~\ref{fig:EuCd2As2:afm:dft} the topological index $\eta_{4I}=2$, which corresponds to an axion insulator. It is stable and independent of almost any direction of the antiferromagnetic vector.
The only exception is given by the alignment of the Eu$^{2+}$ magnetic moments with the $c$-axis, where a crossing is allowed and is also existent~\cite{PhysRevB.98.201116} (see Fig.~\ref{fig:EuCd2As2:afm:dft}~(c)). 
In addition the monoclinic magnetic space group $C_c2/m$, i.e. $\textbf{M}|| 2_{100}$ (or any other secondary axis) introduced above, has the symmetry indicator $\delta_{2m}=1$, describing the difference of mirror Chern numbers at $k_z=0$ and $k_z=\pi$, and, therefore the axion insulator phase is decorated by features of crystalline topology~\cite{elcoro2020magnetic,Ma2020}.

\subsection{\label{sec:afm2} A-type antiferromagnetic ground states under pressure and strain}

Under effects of pressure or strain, both, symmetry changes and electronic structure modifications may influence the topology of \chemform{}.

We first apply hydrostatic pressure preserving all symmetries (see Section~\ref{Computational})
and track the changes of the valence bands in A-type AFM \chemform{} by performing GGA+SOC(in-plane)+U band structure calculations for the configuration with $\textbf{M}||2_{100}$ and $\textbf{M}||m_{100}$.
We restrict the N\'eel vector to inplane directions as these are of most relevant for experimental studies~\cite{PhysRevB.94.045112,PhysRevB.97.214422}.

\begin{figure}[ht]
	\centering
	\includegraphics[width=1.0\linewidth,clip,trim= 0 0 0 0 ]{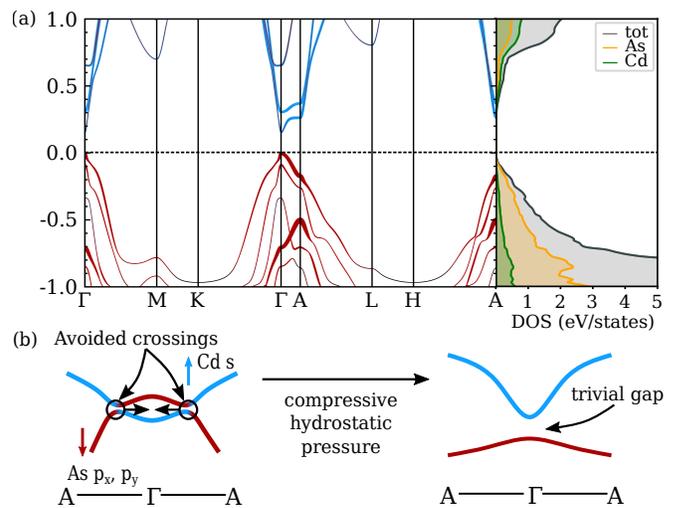}
	\caption{(a) GGA+SOC(in-plane)+U band structure and density of states
	for \chemform{} relaxed under hydrostatic conditions at $p_c\,\approx\,5\,$GPa. The characters of the Cd $s$ and As $p_x$, $p_y$ orbitals are colored in blue and red, respectively.
    (b) Schematic representation of the topological-to-trivial transition along A-$\Gamma$-A when applying compressive hydrostatic pressure to EuCd$_2$As$_2$. }
	\label{fig:EuCd2As2:afm:hydro}
\end{figure}

At $p_c\,\approx\,5$\,GPa, a gap of $\sim 0.15$\,eV opens at the Fermi level in the density of states (see Fig.~\ref{fig:EuCd2As2:afm:hydro}~(a)), and the values of symmetry indicators for the valence bands change from topologically non-trivial to TQC-trivial values.
Equivalently, the coreps of valence bands at maximal $\boldsymbol{k}$-points coincide with those of a linear combination of elementary band corepresentations.
This indicates that, as the applied hydrostatic pressure is increased, the system undergoes a transition to a trivial insulating phase.
This transition is sketched in Fig.~\ref{fig:EuCd2As2:afm:hydro}~(b), and consists on a shift of Cd $s$ and As $p$-orbitals in opposite directions as a results of a change in their hybridization.

Uniaxial strain along the $c$-axis preserves all symmetries of the system similar to the hydrostatic case.
However, it does not cause a change of the band topology for strains we considered of up to $\pm 10\%$.
This straining tunes the exact position of the gapped Dirac cone. At $0\%$ the position of the gapped Dirac cone is about $30$\,meV above the Fermi level, but by applying $-8\%$ to $ -10\%$ it coincides exactly with the Fermi level.
Furthermore, over the whole range of strain $ \pm 10 \%$ the Eu magnetization remains in-plane. 
In contrast, application of shear strain in the A-type antiferromagnet, independently on how the $\gamma$ angle is strained,
leads to a energetic preference of magnetization to be commensurate with the $C_c 2/c$ symmetry over the other symmetries $C_c 2/m$ and $P_S \bar{1}$.

\begin{figure*}[!ht]
		\centering
		\includegraphics[width=0.8\linewidth,clip= true, trim = 0 0 0 0] {figures/fm/bands/FM_electronic_structure_2.pdf}
		\caption{(a)  GGA+U ($U = 5$\,eV) electronic structure (left) and DOS (right) of  \chemform{} in ferromagnetic order. 
        (b) Blow-up of the electronic structure near the Fermi surface along the $\Gamma$-A path. The unit cell is shown on the right.
        (c-e) GGA+SOC+U electronic structure of  \chemform{} in ferromagnetic order along the $\Gamma$-A path by considering 
        the magnetic quantization axis $ \textbf{M}$ (c) along (001), (d) along (100) and (e) along (210).
        The insets of (b-e) denote the corresponding irreducible corepresentations of the magnetic space groups given on the right. Compare with Table~\ref{tab:fm:irreps}. }
	    \label{fig:EuCd2As2:fm:dft}
\end{figure*}

\subsection{\label{sec:fm1} Ferromagnetic ground states}

\begin{table*}[]
\begin{ruledtabular}\begin{tabular}{cccccc}
  $\textbf{M}$ direction            &    space group    & \multicolumn{4}{c}{band corepresentation} \\ \hline
\multirow{4}{*}{$\textbf{M} ||3_{001}^{+}$}  & \multirow{4}{*}{$P \bar{3}m'1$}  &     &  $({}^{1}\!\bar{E} \uparrow P \bar{3}m'1)_{2d}$   &  $({}^{2}\!\bar{E} \uparrow P \bar{3}m'1)_{2d}$   &  $(\bar{E} \uparrow P \bar{3}m'1)_{2d}$  \\\cline{3-6}
&                                                               &$\Gamma$&   $\bar{\Gamma}_5(1) \oplus \bar{\Gamma}_8(1)$  &  $\bar{\Gamma}_6(1) \oplus \bar{\Gamma}_9(1)$   &  $\bar{\Gamma}_4(1) \oplus \bar{\Gamma}_7(1)$  \\
&                                                               &$\Delta$&   $2 \bar{\Delta}_5(1) $  &   $2 \bar{\Delta}_6(1) $  &   $2 \bar{\Delta}_4(1) $ \\
&                                                               &  A     &    $\bar{A}_5(1) \oplus \bar{A}_8(1)$  &  $\bar{A}_6(1) \oplus \bar{A}_9(1)$   &  $\bar{A}_4(1) \oplus \bar{A}_7(1)$   \\ \hline
\multirow{4}{*}{$\textbf{M} ||2_{100}$} & \multirow{4}{*}{$ C2/m $} &    &  $({}^{1}\!\bar{E} \uparrow C2/m)_{4i}$   &   $({}^{2}\!\bar{E} \uparrow C2/m)_{4i}$  &    \\\cline{3-6}
&                                                               &$\Gamma$&   $2 \bar{\Gamma}_4(1) \oplus 2 \bar{\Gamma}_5(1)$   &  $2 \bar{\Gamma}_3(1) \oplus 2 \bar{\Gamma}_6(1)$   &    \\
&                                                               & B      &   $4 \bar{B}_4(1) $  &  $4 \bar{B}_3(1) $   &    \\
&                                                               &   A    &  $2 \bar{A}_4(1) \oplus 2 \bar{A}_5(1)$   &   $2 \bar{A}_3(1) \oplus 2 \bar{A}_6(1)$   &    \\ \hline
\multirow{4}{*}{$\textbf{M} || m_{100} $ or else} & \multirow{4}{*}{$C2'/m'$ or $P \bar{1}$} &     & $(\bar{A} \uparrow C2'/m')_{4i}$ & $(\bar{A} \uparrow P \bar{1})_{2i}$    &        \\\cline{3-6}
&                                                                              &$\Gamma$&   $2 \bar{\Gamma}_2(1) \oplus 2 \bar{\Gamma}_3(1)$  &   $\bar{\Gamma}_2(1) \oplus \bar{\Gamma}_3(1)$  &    \\
&                                                                               &   B/GP    &   $4 \bar{B}_2(1) $  &  $2 \overline{GP}_2(1) $    &    \\
&                                                                               &   A/Z    &   $ 2 \bar{A}_2 (1) \oplus 2 \bar{A}_3 (1)$   &   $ \bar{Z}_2 (1) \oplus \bar{Z}_3 (1)$  &   
\end{tabular}\end{ruledtabular}
\caption{Alignment of the Eu ferromagnetic vector along specific directions and their resulting space groups with the irreducible corepresentations for the given band corepresentation along the symmetry path between $\Gamma$ and A indicated. The notation was adapted from~\cite{Campbell2022}.}
\label{tab:fm:irreps}
\end{table*}

Recently, there has been increased interest on the ferromagnetic phase of EuCd$_2$As$_2$. An important reason for this is the prediction of a single pair of Weyl points just in the vicinity of the Fermi level~\cite{PhysRevB.99.245147}. Ferromagnetic order has been realized in several different ways since then. For instance as a function of band filling~\cite{PhysRevB.101.140402}, by substituting with Ba~\cite{PhysRevB.102.104404}, under hydrostatic pressure~\cite{PhysRevB.104.155124} or by applying a small external magnetic field~\cite{PhysRevB.100.201102}.
Here we investigate the potential of ferromagnetic \chemform{} to exhibit a Weyl semimetallic phase depending on the  direction of Eu$^{2+}$ moments.

In Fig.~\ref{fig:EuCd2As2:fm:dft}~(a-b) we show the GGA+U ($U = 5$\,eV) atom-resolved electronic structure and density of states of FM EuCd$_2$As$_2$ with space group $P\bar{3}m1$. 
The right hand side of Fig.~\ref{fig:EuCd2As2:fm:dft}~(a) contains the band structure, while the left hand side shows the density of states.
Both depend strongly on the magnetization and relate to the strength of the exchange interaction~\cite{PhysRevB.94.045112,PhysRevB.97.214422}.
As in the A-type AFM case, Cd $s$ and As $p_{x}, p_{y} $ orbitals are the dominant ones at the Fermi level, and the density of states near the Fermi level is concentrated along the  $\Gamma$ to A high-symmetry line.

Various interesting topological states emerge upon inclusion of SOC and consideration of different orientations for the moments, as it shown in the GGA+SOC+U calculations in Figs.~\ref{fig:EuCd2As2:fm:dft} (c-e). 
An alignment of the moments parallel to the principal axis (see Fig.~\ref{fig:EuCd2As2:fm:dft} (c) and
Table~\ref{tab:fm:irreps} first row)
preserves $ 3_{001}^{+} $ symmetry, whereas it breaks the unitary 2-fold rotations and mirror reflection.
Hence, the symmetry of the magnetic structure is given by the space group $ P\bar{3}m'1$ (No. 164.89) (see Appendix~\ref{app:symmetry lowering}).
The Wyckoff positions of Cd and As ions are $2d$, and their actual sites are left invariant by site-symmetry groups isomorphic to $3m'$.
With SOC, Cd $s$ transform as the band corepresentation $[({}^1\!\bar{E} \oplus {}^2\!\bar{E})\uparrow G]_{2d}$, whereas the band corepresentation of As $p$-orbitals is $[({}^1\!\bar{E} \oplus {}^2\!\bar{E} \oplus 2\bar{E})\uparrow G]_{2d}$.
These band corepresentations contain more than one corep in the line $\Delta$, which opens the possibility for the band structure to exhibit symmetry-protected Weyl points and band crossings in this high symmetry line, as it is shown in Fig.~\ref{fig:EuCd2As2:fm:dft}~(c)~\cite{Ruan2016}. 

When moments are parallel to a secondary axis, as shown in Fig.\ref{fig:EuCd2As2:fm:dft} (d), 3-fold rotation and rotoinversion symmetries are broken, such that the symmetry is reduced to the magnetic space group $ C2/m$ (No. 12.58).
Cd and As ions sit in WPs $4i$, and their site-symmetry groups are isomorphic to the reflection point group $m$.
This group has two corepresentations which, with SOC, leads to the band corep $[( {}^{1}\!\bar{E} \oplus {}^{2}\!\bar{E}) \uparrow G]_{4i} $.
Since this band corepresentation involves two different coreps, $\bar B_3$ and $\bar B_4$ which do not hybridize on the plane B, band crossings giving rise to Weyl fermions might take place on this plane.

Aligning the magnetic moments with one of the mirror planes (see Fig. \ref{fig:EuCd2As2:fm:dft}(e))
leads to the magnetic space group $ C2'/m'$ (No. 12.62).
Cd and As ions are located in WPs $4i$, whose sites are invariant under site-symmetry groups isomorphic to $m'$.
Moreover, the unitary subgroup of the site symmetry group is given by the identity $1$, therefore, all band corepresentations induced from these positions coincide with $[2\bar{A} \uparrow G ]_{4i}$.
As a consequence, all band corepresentations have the same corepresentation between $\Gamma$ and A, removing any symmetry protection for crossings along the symmetry plane B.
This magnetic space group is compatible with the higher symmetry shear strain case.

For any other direction of moments, or for arbitrary strain modifications, no symmetry other than inversion will be preserved, so the symmetry is lowered to the group $ P\bar{1}$ (No. 2.4), see Fig. \ref{fig:EuCd2As2:fm:dft}~(e).
The Wyckoff position of Cd and As ions is $2i$, with site-symmetry group $1$. 
Just like in the previous case, the band corepresentations induced from this positions involve a single corep on the plane B, preventing the existence of band crossings on its $\boldsymbol{k}$-points.

In contrast to the Dirac points of the AFM case, Weyl points have a local topological protection in $k$-space. Hence, Weyl crossings might take place away from high-symmetry lines and planes. Such crossings might be invisible to symmetry-indicators of topology which stem from corepresentations at maximal $k$-points.

Analogous to the A-type antiferromagnetic case, the index $\eta_{4I}$, describes the topology in systems with inversion $\mathcal{P}$ but without time reversal $\mathcal{T}$ symmetry. We again find $\eta_{4I}=2$ for gapped band structures. In addition, these FM groups might also host mirror-Chern phases labeled by the Chern numbers $z_{2I, i}$, where $i = 1,2,3$ indicates the $k_i = \pi$ planes. It turns out that the values for these Chern numbers are trivial in the gapped FM phases discussed here, and therefore they are axion insulator phases as in the A-type AFM order studied in Sec.~\ref{sec:afm1}. The corresponding topological label is given by $2;0 0 0$ ($ \eta_{4I};z_{2I,1}z_{2I,2}z_{2I,3}$) which corresponds to an axion insulator~\cite{elcoro2020magnetic} as in the A-type AFM order of the previous section.
A special case is again  given by the alignment of the Eu$^{2+}$ magnetic moments with the $c$ axis, where a single pair of Weyl points are possible~\cite{PhysRevB.99.245147}. 
Another special case is the monoclinic magnetic space group $C_c2/m$, i.e. for example $\textbf{M}|| 2_{100}$ introduced above, which has $\delta_{2m}=1$, describing the difference of mirror Chern numbers at $k_z=0$ and $k_z=\pi$, and, therefore, it denotes a topological crystalline insulator~\cite{elcoro2020magnetic}.

\subsection{\label{sec:fm2} Ferromagnetic ground states under strain and pressure}

\begin{figure}[!ht]
	\centering
	\includegraphics[width=1.0\linewidth,clip,trim= 0 0 0 0 ]{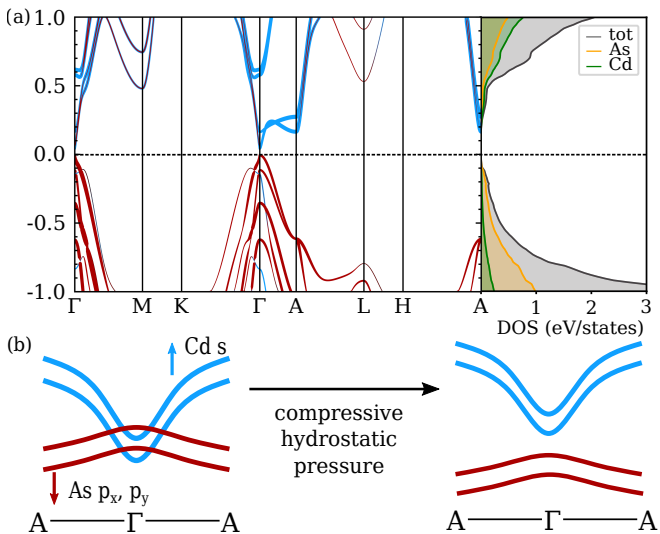}
	\caption{(a) GGA+SOC+U Band structure of \chemform{} 
	 with FM order and Eu$^{2+}$ magnetic moments aligned in $c$-direction at hydrostatic pressure of $p'_c\,\approx\,7.5\,$GPa. Conduction and valence bands do not touch. (b) Schematic representation of applying compressive hydrostatic pressure for the $\Gamma$ to A path. Both subfigures show the Cd $s$ (blue) and As $p_x$, $p_y$ (red) characters.}
	\label{fig:EuCd2As2:fm:hydro}
\end{figure}

Similar to AFM order we apply hydrostatic pressure to FM ordered EuCd$_2$As$_2$.
We focus on the magnetic order pointing along the $c$-direction as this allows to use symmetry indicators to keep track of the fate of the Weyl points. 
We reach a critical pressure, $p'_c\,\approx 7.5\,$GPa where the Weyl points gets annihilated. 
Analogous to the AFM case, this phenomenon can be explained by an overall hybridization change, which forces the bonding and anti-bonding states to move apart from each other. 
This produces an approximately $46\,$meV gap corresponding to a trivial insulating or semiconducting phase without breaking $3_{001}^{+}$ symmetry (see Fig.~\ref{fig:EuCd2As2:fm:hydro}).

Similar to the AFM case, uniaxial strain along the principal-axis does not cause a change of the symmetry indicators for topology where we considered up to $\pm 10\%$.
Furthermore, over the whole range of strain from $ \pm 10 \%$ the Eu magnetization remains in-plane. 
In contrast, application of shear strain in the FM order, independently on how the $\gamma$ angle is strained, leads to a energetic preference of magnetization to be commensurate with the $C 2'/m'$ symmetry over the other symmetries $C 2/m$ and $P \bar{1}$.

\subsection{\label{sec:U} Role of  Hubbard U in the ferromagnetic ground state}

Finally, we investigate the effect of tuning the Hubbard $U$ parameter
on the topology of the electronic structure of ferromagnetic EuCd$_2$As$_2$ . 

\begin{figure}[ht]
		\centering
		\includegraphics[width=\linewidth] {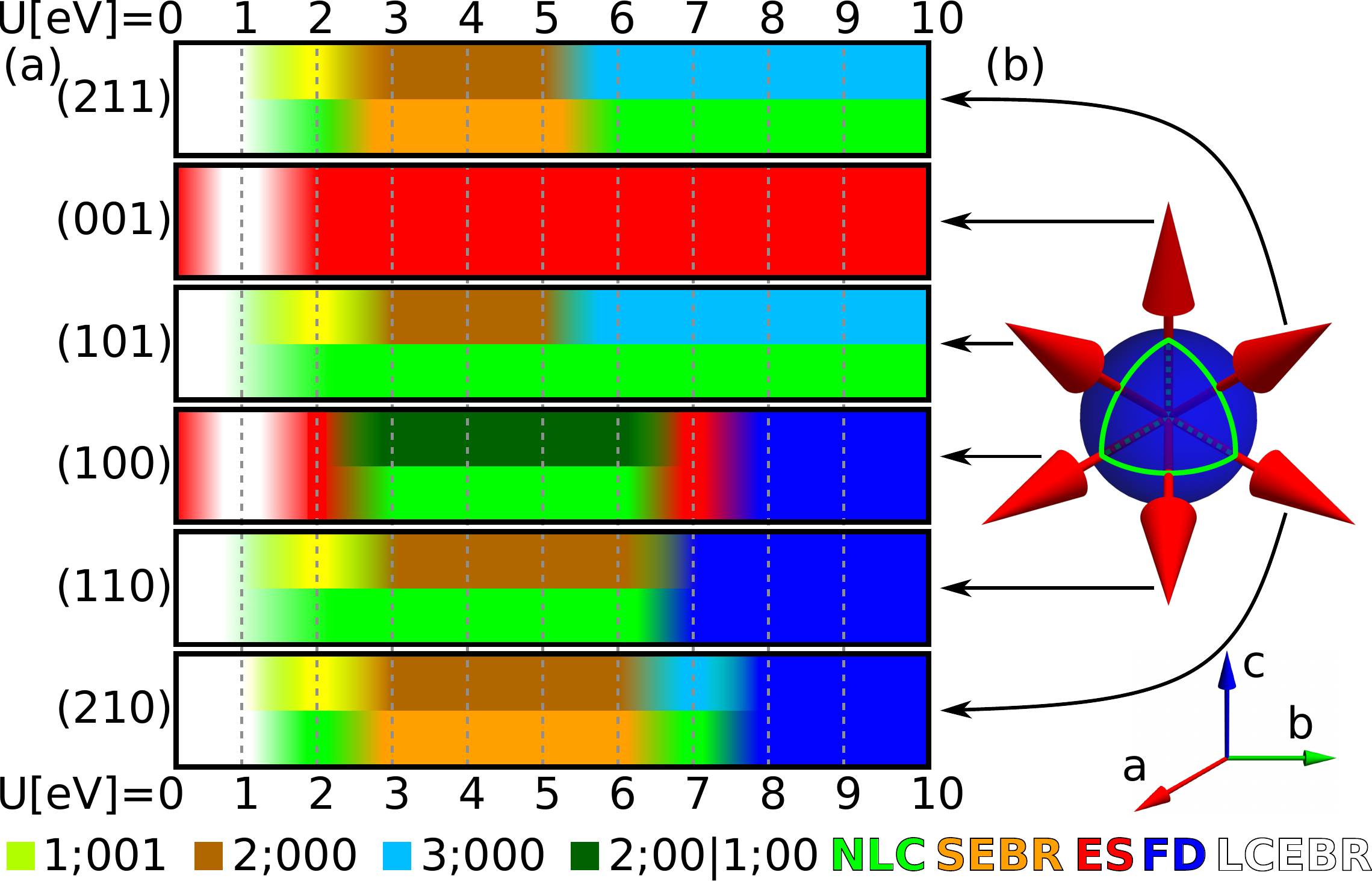}
		\caption{(a) Phase diagram for \chemform{} in dependence of the Hubbard $U$ parameter for different Eu$^{2+}$ magnetization directions of the FM ordered EuCd$_2$As$_2$. 
		From top to bottom we go along the boundary of one eighth of a sphere, (b) visualizes the different direction of the magnetization. The lower colors of the color bars indicate the different types of band corepresentations with respect to the Fermi level, while the upper colors show the topological indices for the topological insulating phases, if applicable, otherwise it is the same as the lower color. The abbreviations were adapted from~\cite{elcoro2020magnetic,xu_high-throughput_2020}, where NLC stands for \textit{No linear combination of EBRs}, SEBR stands for \textit{split EBR}, ES stands for \textit{enforced semimetal}, FD stands for \textit{Fermi degeneracy} and LCEBR stands for \textit{linear combination of EBRs}.}
	    \label{fig:EuCd2As2:fm:phasediag}
\end{figure}
Correlations due to the presence of localized Eu $4f$ states in EuCd$_2$As$_2$ are treated in our analysis within the GGA+U
and GGA+SOC+U exchange-correlation functionals where a Hubbard $U$ term is included in the self-consistent calculations.
In practice, the inclusion of such a term shifts the Eu $4f$ bands to higher binding energies. 
Experimentally it has been observed, that these bands are strongly localized below the Fermi level at about 1\,eV to 2\,eV~\cite{Ma2019}, which corresponds here to $U\,\approx\,5\,$eV.
It is nevertheless interesting to evaluate how the strength of this parameter, and therefore the position of the localized bands affects
the topology~\cite{Ma2020}, even though, for the case of Eu, the 4f bands are strongly localized far below the Fermi level. 
In Fig.~\ref{fig:EuCd2As2:fm:phasediag} we present the result of our analysis by assuming ferromagnetic EuCd$_2$As$_2$ and
different Eu magnetization directions in a range of interaction strengths from $U\,=\,0$ to $10\,$eV.  
According to the elementary band corepresentations (EBRs) for the different spin orientations it can be distinguished between trivial insulators, i. e. linear combinations of EBRs (LCEBRs), accidental Fermi degeneracies (FDs), enforced semimetals (ESs), split EBRs (SEBRs) and no linear combinations (NLCs) of EBRs. 
Furthermore, the topological indices can be calculated for the latter two NLCs and SEBRs. 
The notation was adopted from~\cite{Vergniory2019,xu_high-throughput_2020}. 
The main topological index for this magnetic space group as in all magnetic cases discussed here
is given by $\eta_{4I}$, since time reversal symmetry is broken but inversion is preserved. 
Odd numbers of this index correspond to a Weyl semimetallic phase, while $\eta_{4I}=2$ is an axion insulating phase~\cite{elcoro2020magnetic}. 
For the magnetic space groups $P\bar{3}m'1$, $C2/m$, $C2'/m'$ and $P\bar{1}$ one can further identify $z_{2I,i}$ for $i=1,2,3$ as Chern number in the $k_i=\pi$ planes~\cite{elcoro2020magnetic}.  We will denote these as $\eta_{4I};z_{2I,1}z_{2I,2}z_{2I,3}$. 
The magnetic space group $C2/m$ has an additional mirror index $\delta_{2m}$ describing the difference in the mirror Chern number of $k_z=0$ and $k_z=\pi$, together with $z^{\pm}_{2m,\pi}$ i.~e. the mirror Chern number at $k=\pi$~\cite{elcoro2020magnetic,xu_high-throughput_2020}. We denote these as $\eta_{4I};z_{2I,1}z_{2I,2}|\delta_{2m};z^+_{2m,\pi}z^-_{2m,\pi}$.
As a conclusion we report 4 different topologically insulating non-trivial phases, found by varying the Hubbard $U$ in steps of $1$\,eV parameter and the spin orientations. 
The indices are depicted in Fig. \ref{fig:EuCd2As2:fm:phasediag} and given by $1;001$, $2;000$, $3;000$ and $2;00|1;00$.

\section{\label{sec:concldisc}Conclusion}

In this work we have explored the effects of strain, pressure, magnetic order and spin-orbit coupling on the symmetry and topology of the electronic structure of \chemform{}. 
For that we considered a combination of group theory analysis and DFT calculations.
We not only corroborate previous predictions and observations but we also observe new possible topological phases. 

We found a Dirac semimetallic phase for A-type AFM order and a Weyl semimetallic phase for FM order, in agreement with earlier theoretical predictions~\cite{Ma2020,PhysRevB.99.245147}.
These phases are symmetry protected if $3_{001}^{+}$ rotation symmetry is conserved, otherwise in insulating regimes symmetry indicators were used to identify different topological phases.
We found for instance that \chemform{} is an axion insulator in the absence of the protecting $3_{001}^{+}$ rotation symmetry and a mirror Chern insulator in the presence of $m_{100}$ mirror symmetry.
 
Under application of hydrostatic pressure, we obtain that when the Eu moments are in a A-type AFM order, a band inversion due to the annihilation of gapped Dirac cones occurs, leading to a trivial insulating phase, whereas for ferromagnetically aligned Eu moments the system enters a trivial insulating phase by the annihilation of Weyl points.
In contrast, under strain such as uniaxial strain along the $c$-direction, the topology of the band structure remains robust.
In addition, when we allow for a variable Coulomb repulsion $U$ within GGA+U and GGA+SOC+U , what translates in a shifting in energy of the Eu $4f$ bands, we obtain a rich cascade of topological phases that could be realized in other members of the \chemform{} family by substituting Eu by other magnetic ions where the position of the localized bands are nearer to the Fermi level.

\begin{acknowledgments}
We thank Elena Gati, Paul C. Canfield and Aleksandar Razpopov for discussions and acknowledge support by the Deutsche Forschungsgemeinschaft (DFG, German Research Foundation) for funding through TRR 288 -- 422213477 (project B05) and through QUAST-FOR5249 - 449872909 (project TP4).
The research of R.V. was partially supported by the National Science Foundation under Grant No. NSF PHY-1748958 during a visit to the Kavli Institute for Theoretical Physics (KITP), UC Santa Barbara, USA for participating in the program “A Quantum Universe in a Crystal: Symmetry and Topology across the Correlation Spectrum”.
\end{acknowledgments}

\appendix

 \section{\label{app:gs:table}Ground state relaxation}
 \begin{table}[!h]
\begin{tabular}{|c|r|r|}
\hline
\chemform            & \multicolumn{1}{c|}{experiment~\cite{zaac.201100179}} & \multicolumn{1}{c|}{theory} \\ \hline
a [\r{A}/f.u.]       & 4.450                                                 & 4.508                       \\ \hline
c [\r{A}/f.u.]       & 7.350                                                 & 7.403                       \\ \hline
Vol [\r{A}$^3$/f.u.] & 145.542                                               & 150.212                     \\ \hline
Cd $z$               & 0.247                                                 & 0.246                       \\ \hline
As $z$               & 0.633                                                 & 0.634                       \\ \hline
\end{tabular}
\caption{
Unit cell parameters, volume and $z$ fractional coordinates for Cd and As of \chemform{} in the $P\bar{3}m1$ space group from powder X-ray diffraction measurements~\cite{zaac.201100179} and
after DFT relaxation as explained in the text. The Wyckoff positions are $1a$ for Eu and $2d$ for Cd and As . 
}
\label{tab:unitcell}
\end{table}

\section{Magnetic structures and their magnetic groups} \label{app:symmetry lowering}

In this section we discuss the way each of the magnetic configurations investigated in the main text lowers the symmetry of the system. We consider as a starting point the symmetry of the non-magnetic ionic lattice, whose symmetry is given by the space group $P\bar{3}m11' = P\bar{3}m1 \otimes \mathcal{T}$ (No. 164.86, with time reversal $\mathcal{T}$). Its symmetry elements that are not equivalent up to translations are given by
\begin{equation}
\lbrace 1, 3_{001}^{\pm}, 2_{100}, 2_{010}, 2_{110}, \bar{1}, \bar{3}_{001}^{\pm}, m_{100}, m_{010}, m_{110} \rbrace \otimes \mathcal{T}.
\end{equation}
Formally speaking, these form a complete set of coset representatives for a coset decomposition with respect to the translation group. For the sake of simplicity we will denote symmetry elements by their point group operation instead of using the Seitz notation and we will only consider the coset representatives for the following symmetry groups. As discussed in Section~\ref{sec:nm}, the application of strain can reduce the symmetry of the system in different ways, depending on the type of strain and its direction. Shear strain acting e.g. in [100]-direction will reduce the symmetry to the space group $C2/m1' = C2/m \otimes \mathcal{T}$ (No. 12.59) with symmetry elements $\lbrace 1, 2_{100}, \bar{1}, m_{010} \rbrace \otimes \mathcal{T}$, whereas any other strain will either leave all the symmetries intact or reduce the to space group to $P\bar{1}1' = P\bar{1} \otimes \mathcal{T}$ (No. 2.5) with symmetry elements $\lbrace 1, \bar{1} \rbrace \otimes \mathcal{T}$.

Magnetically ordered structures do not contain $\mathcal{T}$ as a symmetry element and
are described by magnetic space groups which might contain $\mathcal{T}$ combined with crystal symmetries other than identity. Symmetry operations also act on the magnetization which transforms as a pseudovector. Due to the spontaneous breaking of translation symmetry in $c$-direction, all antiferromagnetic structures contain the symmetry $\mathcal{T}' = \mathcal{T} \tau$, where $\tau = (0, 0, 1/2)^{T}$ is half a translation in $c$-direction of the primitive magnetic unit cell. They are thus all type IV Shubnikov groups. Depending on whether the magnetization is parallel to the principal axis, one of the three secondary axes or a mirror plane, one obtains the space groups $P_c \bar{3}c1$ (No. 165.96), $C_c 2/m$ (No. 12.63) and $C_c 2/c$ (No. 15.90), while for ferromagnetic ordering one finds $P\bar{3}m'1$ (No. 164.89, type III), $C2/m$ (No. 12.58, type I) and $C2'/m'$ (No. 12.62, type III). Their symmetry elements are listed in Table~\ref{tab:symmetry_elements}.

\begin{table*}[]
\begin{ruledtabular}\begin{tabular}{cccc}
Magn. order & $\textbf{M}$ direction & space group & symmetry elements  \\ \hline
AFM & $\textbf{M}||3_{001}^{+}$ & $P_c \bar{3} c 1$ & \vtop{\hbox{\strut $\lbrace 1, 3_{001}^{\pm}, \tau 2_{100}, \tau 2_{110}, \tau 2_{010}, \bar{1}, \bar{3}_{001}^{\pm}, \tau m_{100}, \tau m_{110}, \tau m_{010},$}\hbox{\strut $\mathcal{T}', \mathcal{T}' 3_{001}^{\pm}, \mathcal{T} 2_{100}, \mathcal{T} 2_{110}, \mathcal{T} 2_{010}, \mathcal{T}' \bar{1}, \mathcal{T}' \bar{3}_{001}^{\pm}, \mathcal{T} m_{100}, \mathcal{T} m_{110}, \mathcal{T} m_{010} \rbrace$}} \\ \hline
AFM & $\textbf{M}||2_{100}^{+}$ & $C_c 2/m$ & $\lbrace 1, 2_{100}, \bar{1}, m_{100}\rbrace \otimes \mathcal{T}'$ \\ \hline
AFM & $\textbf{M}||m_{100}^{+}$ & $C_c 2/c$ & $\lbrace 1, \tau 2_{100}, \bar{1}, \tau m_{100}, \mathcal{T}', \mathcal{T} 2_{100}, \mathcal{T}' \bar{1}, \mathcal{T} m_{100} \rbrace$ \\ \hline
FM & $\textbf{M}||3_{001}^{+}$ & $P \bar{3} m' 1$ & $\lbrace 1, 3_{001}^{\pm}, \bar{1}, \bar{3}_{001}^{\pm}, \mathcal{T} 2_{100}, \mathcal{T} 2_{110}, \mathcal{T} 2_{010}, \mathcal{T} m_{100}, \mathcal{T} m_{110}, \mathcal{T} m_{010} \rbrace$ \\ \hline
FM & $\textbf{M}||2_{100}^{+}$ & $C2/m$ & $\lbrace 1, 2_{100}, \bar{1}, m_{100} \rbrace$ \\ \hline
FM & $\textbf{M}||m_{100}^{+}$ & $C2'/m'$ & $\lbrace 1, \bar{1}, \mathcal{T} 2_{100}, \mathcal{T} m_{100} \rbrace$
\end{tabular}\end{ruledtabular}
\caption{Coset representatives with respect to the translation group of all of the space groups under consideration depending on the orientation of the magnetization.}
\label{tab:symmetry_elements}
\end{table*}

\section{\label{app:afm:Ushift}Magnetic moments of Eu, Cd and As}
\begin{figure}[]
		\centering
		\includegraphics[width=\linewidth] {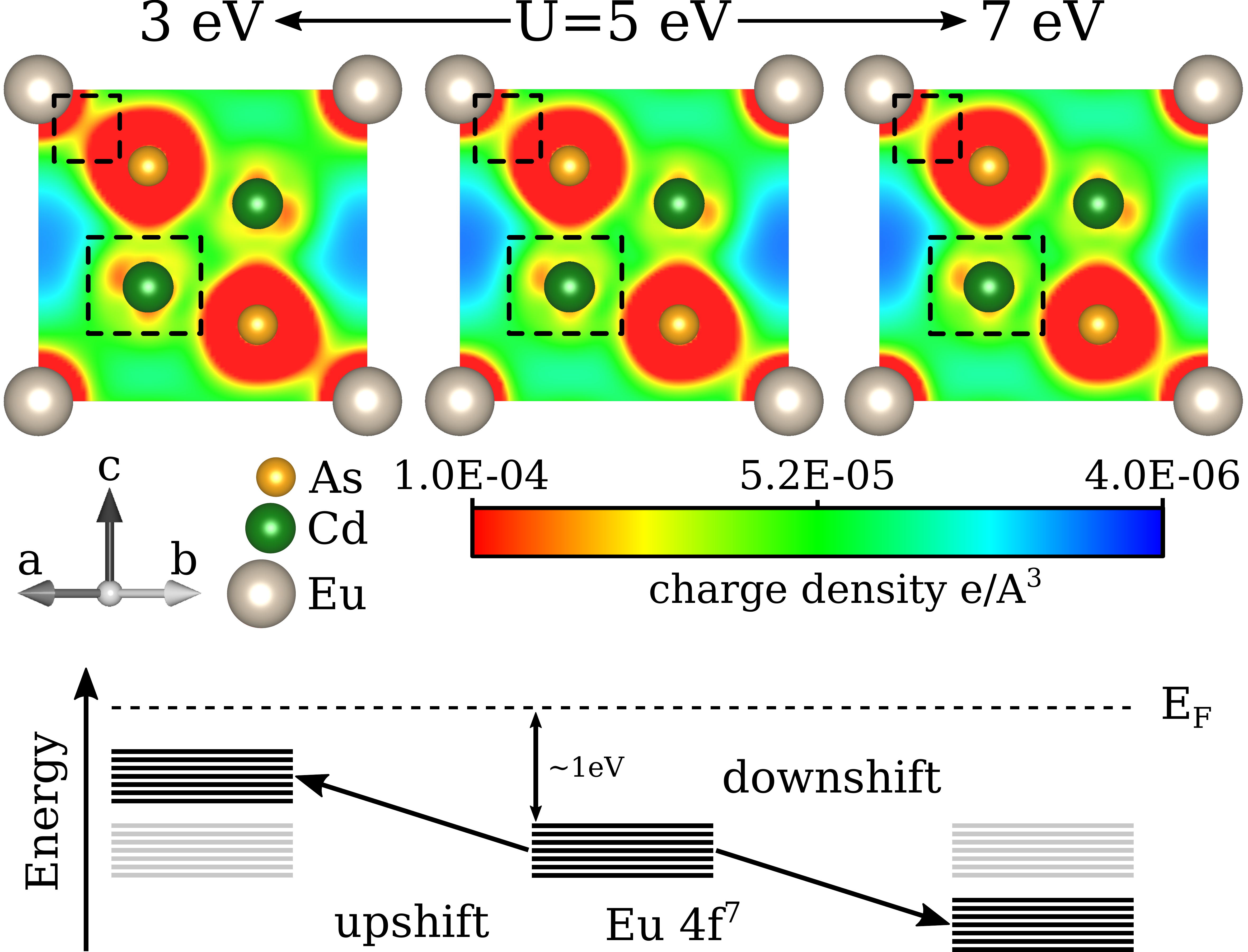}
		\caption{
        The top panel shows charge density plots in dependence of the $U$ parameter for the values $3$\,eV, $5$\,eV, and $7$\,eV for the AFM case.
        The charge densities are plotted onto one of the mirror planes along the $c$-direction between $(0, 0, 0)^{T}$ and $(0,0,1/2)^{T}$.
        The dashed rectangles emphasize the regions with changes compared to $U = 5$\,eV.
        The lower panel shows a schematic figure for the effect of varying the $U$ parameter.
        }
	    \label{fig:EuCd2As2:afm:U}
\end{figure}

For the electronic structure calculations, the magnetism of Eu influences Cd and As, which usually do not carry any magnetism,
 even if 
the Eu 4f electrons are strongly localized at about ($\approx -1$\,eV) from the Fermi level.
For our calculations the Eu magnetic moment converges to a value of $6.916\,\mu_B$. 
Cd has the lowest magnetic moment of $0.005\,\mu_B$ and As is approximately a factor of three stronger with $0.014\,\mu_B$.
The two induced magnetic moments are small, however, they are big enough to reflect the symmetry change due to the variation of the Eu moment direction as one can see in Fig.~\ref{fig:EuCd2As2:afm:dft}.
To increase or decrease the influence of the Eu moments, we varied the position of the Eu 4f electrons via the Hubbard $U$ parameter, see Fig.~\ref{fig:EuCd2As2:afm:U}.
We picked three different U values $3$\,eV, $5$\,eV, and $7$\,eV and visualized the charge densities $\rho=\rho_{\uparrow}+\rho_{\downarrow}$.
As shown in the figure, for increasing $U$, which corresponds to a shift farther away from the Fermi level (as described in Fig.~\ref{fig:EuCd2As2:afm:U} lower panel), Eu has less effect on the shape of the charge densities around Cd and As, see Fig.~\ref{fig:EuCd2As2:afm:U} upper panel highlighted by the dashed rectangles.
A decreased $U$ has exactly the opposite effect.
The calculated magnetic moment of Eu is constantly increasing in steps of $0.03\,\mu_B$ from $3$\,eV to $7$\,eV, As shows the same behaviour with less magnitude of about $3\,\cdot\,10^{-4}\,\mu_B$.
Cd on the other hand behaves opposite, it increases in steps  of $4.5\,\cdot\,10^{-4}\,\mu_B$. This may be attributed to the location of the Cd 5s bondig states in the energy range of $-4$\,eV to $-5$\,eV, which then form a stronger hybridization with the Eu 4f states, compare Fig.~\ref{fig:EuCd2As2:afm:U}.
By removing the Eu 4f states completely, the magnetization of Cd and As reduces to 0, as expected.

\section{\label{app:bandrep} Short introduction to band (co)representations}
Central to the framework of topological quantum chemistry is the notion of band representations~\cite{Zak1980,Cano_2018}, according to which Bloch wave functions in a set of bands inherit their symmetry properties, protected by a space group $G$, from the localized orbitals they originate from.
In real space an orbital is exponentially localized at some point in the unit cell.
For the symmetry analysis, it only matters which Wyckoff position it corresponds to.
These orbitals sit in Wyckoff positions, and their transformation under symmetries is described by a representation $\rho$ of the corresponding site-symmetry group.
A band representation is a representation of the space group which acts on a set of localized orbitals and which respects their symmetry behavior characterized by $\rho$.
In the language of group theory one says that the band representation $\rho_G$ is induced from $\rho$, formally denoted by $\rho \uparrow G$.
Since every site symmetry group is isomorphic to a point group, it is common to use a representation of the point group for $\rho$. 
Different Wyckoff positions may have isomorphic site symmetry groups, so to uniquely define the band representation one writes $(\bar{\rho} \uparrow G)_{\text{Wyk}}$, where $\bar{\rho}$ is the representation of the point group and Wyk is the Wyckoff position.

If the Wannier functions of a band (or set of bands) are exponentially localized, they will transform like orbitals, hence its symmetry behaviour can be described by a band representation $\rho_{G}$.
The representations under which the band(s) at high symmetry points $\vec{k}$ in reciprocal space transform, characterize the band representation and are used as symmetry indicators.
They are given by subducing the band representation to the little co-groups $G_{\vec{k}}$ of high symmetry points, i.e. restricting the symmetry operations of the space group $G$ to those that leave $\vec{k}$ invariant (where elements related by translation are treated equivalent, so that the little co-group is isomorphic to a finite point group), formally denoted as $\rho_{G} \downarrow G_k$.
If the representations of the little co-groups of some band at high symmetry points cannot be related to any possible band representation, its Wannier functions cannot be exponentially localized indicating non-trivial topology.
A variety of different topological phases, such as semimetals may also be diagnosed by symmetry indicators.

For magnetic space groups representation theory has to be modified to account for antiunitary symmetry elements. Instead of representations for unitary groups one works with corepresentations, which are constructed form ordinary representations using the Frobenius–Schur indicator. Band corepresentations induced from magnetic point groups behave similarly to band representations in that one can form a direct sum of elementary corepresentations to obtain composite band corepresentations that describe localized atomic orbitals.

\newpage

\bibliography{apssamp}

\end{document}